**Title:** Fetus dose calculation during proton therapy of pregnant phantoms using MCNPX and MCNP6.2 codes

Authors: Marijke De Saint-Hubert[a], Katarzyna Tymińska[b], Liliana Stolarczyk[c,d], Hrvoje Brkić[e,f] *

[a] Belgian Nuclear Research Centre (SCK CEN), Boeretang 200, BE-2400 Mol, Belgium
[b] National Centre for Nuclear Research, A. Sołtana 7, Otwock 05-400, Poland
[c] Danish Centre for Particle Therapy – DCPT, Denmark
[d] Cyclotron Centre Bronowice - CCB IFJ PAN, Poland
[e] Department of biophysics and radiology, Faculty of medicine, J. J. Strossmayer University of Osijek, J. Huttlera 4, HR-31000 Osijek
[f] Department of biophysics, biology and chemistry, Faculty of dental medicine and health, J. J. Strossmayer University of Osijek, Crkvena 21, HR-31000 Osijek

*hbrkic@mefos.hr, J. Huttlera 4, HR-31000 Osijek, CROATIA – corresponding author

**Abstract**

Radiotherapy of pregnant cancer patients is not common, but when applied accurate assessment of the fetus dose is required, especially since the treatment planning systems are not able and do not allow accurate assessment of out-of-field doses. Proton therapy significantly reduces out-of-field doses, such as dose to the fetus, when compared to the photon radiotherapy techniques and as such could be promising for pregnant cancer patients.
Within this study Monte Carlo calculations are performed on the three different computational phantoms representing pregnant women, all in second trimester of pregnancy. Simplified proton beam to the pregnant women brain was modelled, and the total dose equivalent (normalized per target dose) to the fetus was calculated.
Between MCNPX and MCNP6.2 code versions we observed up to 6% difference. In this work 3 groups participated and the variation between the groups was 8% and 12% for MCNPX and MCNP6.2, respectively. Depending on the phantom used in the calculations the fetal dose was between 0.4 and 0.8 µSv/Gy. Major contribution to the total dose came from the neutrons, with only 20% of the dose coming from photons, while other particles have negligible contributions. Results of the total dose equivalent differ by factor of two, when different phantoms were used, due to geometrical and anatomical characteristics of mother and fetus position.

**Highlights**

- Code versions agreed within 6% in the fetus absorbed dose calculation
- Variation between the groups was 8% and 12% for MCNPX and MCNP6.2
- Fetal dose was between 0.4-0.8 µSv/Gy due to different geometries between phantoms
- No impact was observed for varying female and fetus tissue composition

# Introduction

Cancer during the pregnancy is not common since it is diagnosed in less than 0.1% of all diagnosed cancers (Donegan, 1983; Kal and Struikmans, 2005; Pavlidis, 2002). However, due to increase of the average age of woman in their first pregnancy, this prevalence might increase (Kal and Struikmans, 2005). Breast and brain cancer as well as Hodgkin's disease are the most frequent tumors occurring during pregnancy. On average 70 % of diagnosed cancers during pregnancy are treated, but only 3 % receive radiotherapy. Nevertheless, this type of treatment could be an alternative to chemotherapy especially during the first trimester (Kal and Struikmans, 2005). Still a significant number of radiotherapy centers do not have experience in treating pregnant patients.

Over the past few decades radiotherapy made some significant technological advances to improve the effectiveness and tolerability of the treatment (Mazzola et al., 2019). The number of the pregnant patients that undergo such treatments is not high, so there is no fundament for any clinical recommendation (Mazzola et al., 2019). The most frequently used threshold (100 mGy) for the dose to the fetus is the one recommended by ICRP (Streffer et al., 2003).

As fetus is located out of the radiation field, treatment planning systems (TPS) is not able to accurately estimate the dose to fetus (Shine et al., 2019). Therefore, it is crucial to perform measurements or calculations of fetus doses. Previous measurements during photon radiotherapy have shown that out-of-field doses might arise from radiation scattered within the patient, radiation scattered from the collimators and leakage from the treatment head that mainly depends on the treatment device. Proton radiotherapy has many advantages compared to photon therapy, i.e. there is decreased dose at the entrance to the tissue due the slowdown of the protons and there is no exit dose. Some studies showed that there is a tenfold reduction to the fetus dose when protons are used (Geng et al., 2015). Nevertheless the dosimetry is challenged by the creation of secondary neutrons leading to a mixed field of out-of-field radiation comprising of protons, neutrons, photons and alphas with an increased and variable biological effectiveness.

Monte Carlo simulations of radiation transport are commonly used by medical physicist to obtain accurate information that cannot be measured or evaluated by TPS. Moreover, MC allows researchers to obtain neutron spectra for correction of detector response or to obtain out-of-field doses during radiotherapy (Stolarczyk et al., 2018; Wochnik et al., 2020). However, they may also serve as a golden standard in many other medical applications (Kolacio et al., 2021; Rabus et al., 2021). In the computational dosimetry, voxelized human phantoms are commonly used to either assess the doses to specific organs, or to provide effective dose (Zaidi and Xu, 2007). Computational resources development allowed researchers to produce more accurate models of the human body and several voxelized phantoms of pregnant women are available (Becker et al., 2008; Dimbylow, 2006; Maynard et al., 2014).

The aim of this study is to calculate the fetus dose during proton brain therapy using Monte Carlo (MC) calculations on three different voxelized phantoms of pregnant women and compare several independent simulations performed by 3 different groups and different versions of MCNP (MCNPX 2.7.0 and MCNP6.2).

## Materials and methods

Three groups from three different institutes, approached this task. The groups are marked as 1, 2 and 3, in order not to present their results directly.

### MCNP

All the participants used Monte Carlo N-Particle transport code (MCNP), but there were differences in the versions. Eventually all participants agreed to use both MCNPX 2.7.0 (Pelowitz, 2011) and MCNP6.2 (Werner et al., 2018b, 2018a) versions of the code to spot the differences between the codes, participants and approaches. One major difference between these two codes is that MCNPX uses Bertini Intranuclear Cascade (INC) model and Dresner pre-equilibrium model as the default (Bertini, 1969, 1963), while MCNP6.2 uses the Cascade-Exciton Model CEM03.03 (Mashnik et al., 2008, 2005). Also MCNP6.2 allows performing the calculations on multiple cores which, together with skipping the geometry check (that is default in MCNPX), speeds up the calculations significantly.

### Phantoms

Three different computational pregnant female phantoms were chosen for the simulations, one developed by Helmholtz University Zentum Münich, named Katja (Becker et al., 2008), and two from the University of Florida (UF) family of phantoms at different gestational stages (Maynard et al., 2014) (Figure 1). Katja phantom represents pregnant women at 24 weeks of pregnancy, and UF phantoms have more stages, that are measured as post conceptual and only two stages of UF phantoms were used in this study. The UF phantoms are referred as UF20 (for 20$^{th}$ week post conception) and UF25 (for 25$^{th}$ week post conception) in the further text. If the post conceptual weeks are recalculated to the pregnancy weeks, two weeks should be added to the UF phantoms in order to speak in the same terms. Characteristics of the computational phantoms are given in the Table 1. All the phantoms have similar geometrical characteristics, Katja is slightly taller (4 cm) than UF phantoms, the masses of the mother are also similar (within 2 kg) and Katja fetus mass is in between of UF20 and UF25.

UF phantoms are built from voxel lattices of 2 sizes – the ones that represent the fetus are smaller, allowing the modeling of more details. Therefore, the number of voxels in UF phantoms is significantly higher when compared to the Katja phantom, built of same voxel size.

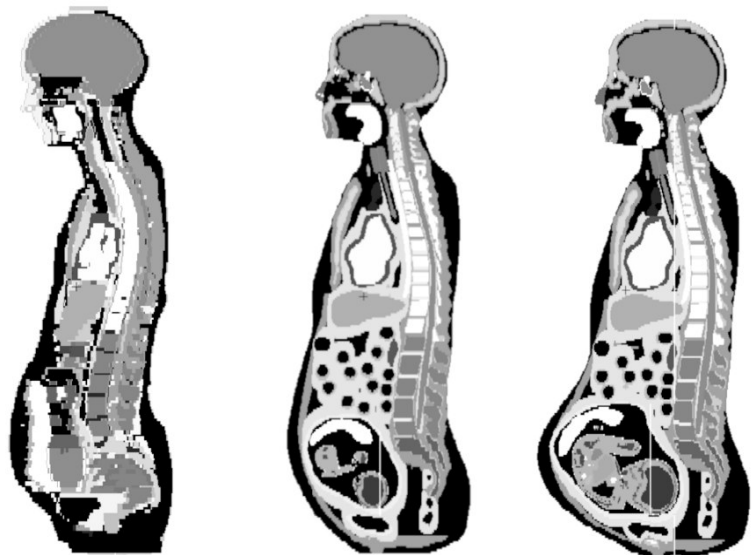

Figure 1. Representations of phantoms: Katja (left), UF20 (middle) and UF25 phantom (right).

Table 1. Physical and geometrical characteristics of the different phantoms used.

| Phantom | Katja | UF20 | UF25 |
|---|---|---|---|
| Weeks of pregnancy | 24 | 22 | 27 |
| Female height [cm] | 168 | 164 | 164 |
| Female mass [kg] | 63.6 | 63.6 | 65.8 |
| Fetus mass [g] | 730 | 468 | 986 |
| Voxel size [mm$^3$] | 1.775x1.775x4.84 | 1.26x1.26x2.7 Mother<br>0.301x0.301x0.301 Fetus | 1.26x1.26x2.7 Mother<br>0.381x0.381x0.381 Fetus |
| Number of voxels (*10$^6$) | 15.7 | 53.65 fetus<br>57.24 Mother | 51.96 fetus<br>66.78 Mother |

MC simulation conditions

Proton beam

For the purpose of comparison, a simplified proton beam that targets the center of the brain was chosen. The beam was coming from the right side of the phantom head, it was circular with 3 cm diameter and Spread Out Bragg Peak (SOBP) with a range of 10 cm and modulation of 5 cm in water. In total 21 proton energies were modelled ranging from 78 to 116 MeV with a Gaussian energy distribution (sigma 1.5%) and appropriate weightings as presented in the supplemental Table 1.

Brain target dose

To display an SOBP profile inside the brain a mesh tally was used providing results in MeV/cm$^3$/source proton. Target dose (MeV/g/proton) was obtained by correcting for the brain density. The out-of-field dose quantities were normalized to the averaged target dose/proton form the SOBP plateau to derive dose quantities per target dose.

Fetus dose

As a variance reduction technique, geometry splitting was applied, namely all the particles entering the uterus have importance 4, while for those particles entering any of the fetus organs this was further increased to 16. The absorbed dose from protons, neutrons, photons, electrons and alphas for whole fetus was calculated using +F6 tally, while the neutron spectra were recorded using F4 tally. All simulations were conducted for at least $10^8$ particles, and all tallies had relative error below 10%.
Since the simulations are performed in mixed field, to distinguish the contribution from different particles, phase space (PHSP) files were created, where only one type of particle was recorded. Surface for recording the particles was a cylinder placed coaxial from the source cylinder with radius of 10 cm. The simulations that started from each phase space file enabled us to determine contribution to the fetus dose from each particle type.

Several sets of simulations, to test the influence of different codes, phantoms and material compositions, were conducted as represented in Table 2. In all simulation sets the atomic number of the element, mass number of the nuclide and cross-section identifier were selected for each material

(Supplemental Table 2) from available data libraries (Collins, 2014). Photonuclear and proton interactions were selected using material card nuclide substitution (MX card) for carbon and iron in all defined materials for simulation sets 1, 4 and 5.

Table 2. Overview of simulation sets presenting phantom geometry, mother and fetus material composition

| Set | Phantom geometry | Mother material | Fetus material |
|-----|------------------|-----------------|----------------|
| 1 | Katja | Katja | Katja |
| 2 | Katja | Katja | UF25 |
| 3 | Katja | UF25 | UF25 |
| 4 | UF25 | UF25 | UF25 |
| 5 | UF20 | UF20 | UF20 |

Calculation of fetus dose equivalent

Fetus dose equivalent was calculated summing 2 separate contributors:
1) Neutron dose equivalent obtained as explained in Romero-Expósito et al (2016). Assuming the validity of the kerma approximation (i.e. secondary charged particle equilibrium exists), the simulated fluence data (F4:n) are used together with the kerma factors *k(E) and* quality factor as a function of neutron energy (*Q(E)*) to derive the neutron dose equivalent using the following equation:

$$H = \Phi \int_E Q(E) \cdot k(E) \cdot \frac{d\varphi_i(E)}{dE} \cdot dE \qquad (1)$$

where $\Phi$ is the total neutron fluence and $\frac{d\varphi_i(E)}{dE}$, the energy spectrum of the unit neutron fluence.
2) Photon dose equivalent calculated from the absorbed dose due to secondary photons (F6:p) and using the quality factor of photons ($Q_\gamma=1$).

Both dose quantities were normalized to the target dose, obtained as average at the SOBP plateau, to derive a fetus dose (equivalent) per target dose expressed as nGy/Gy or nSv/Gy.

## Results

Brain target doses

In Figure 2 the proton beam hitting the brain of the computational phantom is shown as well as the SOBP profile. Both phantoms (Katja and UF25) show on average of 0.42 MeV/g/source proton at the SOBP plateau. The elevated dose in the plateau of the SOBP (at ±1.5cm) is related to the increased density of the cranium when entering the brain of the patient.

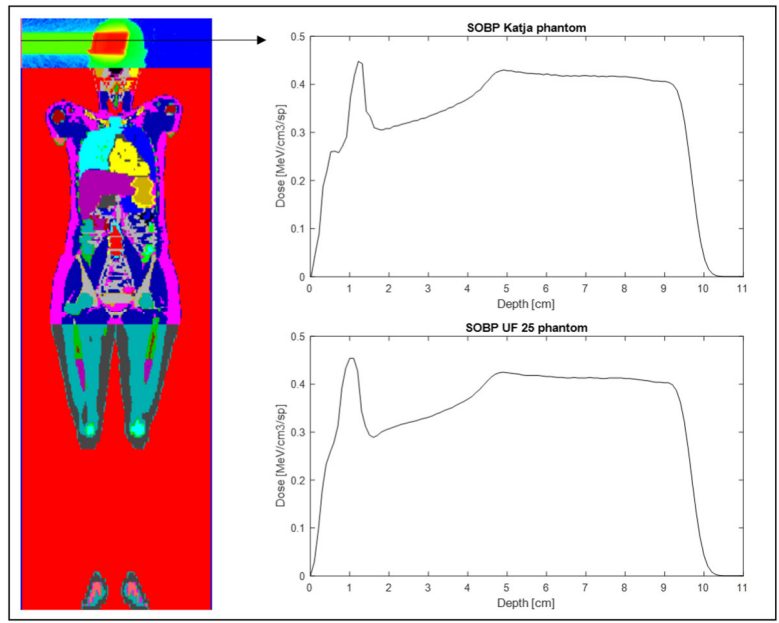

Figure 2. On the left the voxelized Katja phantom with mesh tally in the brain representing the entrance of the SOBP in the brain. Corresponding SOBP profiles are shown on the right for Katja phantom as well as for UF25 phantom, as obtained by group 1. The depth starts from the entrance of the beam in the patient head.

Variation between participants and codes

The absorbed dose in fetus calculated with two different codes by three different groups for the Katja phantom are presented in Table 3.
The variation in the fetus total dose between the groups is 8% for MCNPX while it was slightly higher for MCNP6.2 and variation reached up to 12%. In general, photon/electron doses match better between the groups for MCNPX and MCNP6.2 (within 8% and 6 % respectively) compared to the dose deposited by protons, which is more variable reaching up to 21% and 26% respectively.
This difference may be attributed to slightly different definition of the beam parameters and positioning of the beam to the center of the brain between participants.
On the other hand when running the same input file MCNPX and MCNP6.2 differences are within 6% for the total absorbed dose. This can be expected as the same cross section libraries were selected and the different default nuclear models between MCNPX and MCNP6.2 will not play a major as the proton source energy is up to only 116 MeV and nuclear models are only used above 150 MeV in the MCNP code.

Nevertheless, the difference was somewhat higher for proton dose and we noted that the difference reached up to 13% between MCNPX and MCNP6.2 for group 3 which was unexpected.

Table 3. Absorbed dose in fetus normalized per target dose [nGy/Gy] calculated for different particles by different groups and codes calculated on the Katja phantom (corresponding to scenario 1). Differences between MCNPX and MCNP6.2 are shown as well as variation between the groups for the different particle type.

| Group | Particle type | MCNPX | MCNP6.2 | Difference (%) |
|---|---|---|---|---|
| 1 | Proton | 30 | 30 | 0% |
| 1 | Neutron | 42 | 42 | 0% |
| 1 | photon/electron | 93 | 92 | 1% |
| 1 | All ("+f6") | 165 | 164 | 1% |
| 2 | Proton | 22 | 23 | -4% |
| 2 | Neutron | 42 | 41 | 2% |
| 2 | photon/electron | 104 | 101 | 3% |
| 2 | All ("+f6") | 168 | 165 | 2% |
| 3 | Proton | 34 | 39 | -13% |
| 3 | Neutron | 53 | 55 | -4% |
| 3 | photon/electron | 103 | 108 | -5% |
| 3 | All ("+f6") | 190 | 202 | -6% |
| Variation | Proton | 21% | 26% | |
| Variation | Neutron | 14% | 17% | |
| Variation | photon/electron | 6% | 8% | |
| Variation | All ("+f6") | 8% | 12% | |

Total dose equivalent - contribution from different particles

In order to calculate the total dose equivalent per target dose we used phase space files to assess the dose from each particle type individually. We created a cylindrical surface around the beam axis and recorded various particles (neutrons, protons, photons and electrons) crossing the surface in separate phase space files. In subsequent calculations the files were used as source input to calculate the dose contribution to fetus from each particle type. Both neutrons and photons created within the brain of the patient contribute to the fetus dose while protons and electrons do not reach the fetus (Table 4). Overall, the contribution from photons is around 20% and mostly neutrons will contribute to the total dose equivalent. As such all proton doses measured inside the fetus (Table 3) are attributed to recoil protons created by neutrons.

This confirmed the total dose equivalent in the fetus consists of photons and neutrons and the calculation defined in materials and methods can be used to calculate the total dose equivalent per target dose.

Since the results of all institutions show the similar trend, only results from one institution are shown.

Table 4. Absorbed dose contribution from each particle determined by PHSP file. Results from the group 2 for the Katja phantom are shown

|  | Neutron phase space | Proton phase space | Photon phase space | Electron phase space | Total |
|---|---|---|---|---|---|
| Dose equivalent per target dose [nS/Gy] | 615 | 0.00001 | 144 | 0.0001 | 759 |
| Contribution to total dose equivalent | 81 % | 0 % | 19 % | 0 % | 100 % |

Total dose equivalent - differences between phantoms

Simulations with UF phantoms at two gestational ages were performed. Photon dose, neutron dose equivalent and total dose equivalent (all normalized per target dose) were determined for each phantom and the results are summarized in Table 5. Since the results from all the institutions show similar trend only results from Group 3 are shown. To our surprise the total dose equivalent for Katja phantom was almost two times higher than the total dose equivalent determined for the UF phantoms and similar differences are observed for photons and simulated neutron dose data. Between UF20 and 25 weeks the difference is small, with a slight elevation in the UF25 weeks compared to UF20 weeks which is expected as the fetus is larger and will get closer to the target area.

Table 5. Results of total dose equivalent in different phantoms using MCNP6.2 code. Results from scenarios 1, 4 and 5 and Group 3 are shown.

|  | Dose quantities | | | Difference to Katja (%) | |
|---|---|---|---|---|---|
|  | Katja | UF20 | UF25 | UF20 | UF25 |
| Photon dose per target dose [nGy/Gy] | 108 | 60 | 64 | 44% | 40% |
| Neutron dose equivalent per target dose [nSv/Gy] | 672 | 295 | 332 | 56% | 51% |
| Total dose equivalent per target dose [nSv/Gy] | 780 | 355 | 396 | 54% | 49% |

In order to understand the differences observed between the phantoms we modelled the total absorbed dose in a mesh tally as is shown in the Figure 3. This figure shows that the dose in the belly of the Katja is elevated compared to the UF 25 phantom.

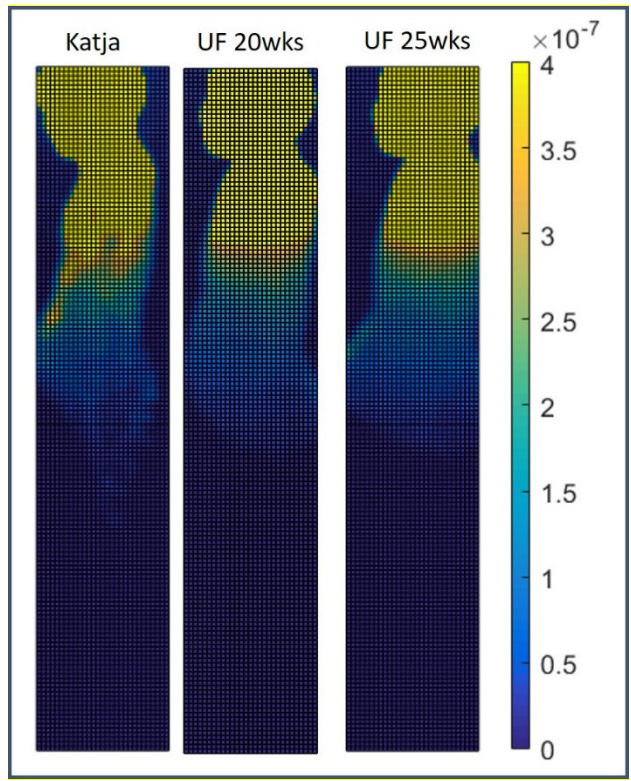

Figure 3. Image of the mesh tally the total absorbed dose through the phantom. Results from Group 1 are shown for Katja, UF20 and UF25.

Impact of tissue composition

We noted differences in the definition of the tissue composition between Katja and UF20/25 for both female and fetus. Therefore, we tested also the influence of the material composition, which showed that there is no significant difference after modifying the chemical tissue compositions of the Katja phantom. Among the groups and codes, it does not differ more than 0.3 %, which is within the statistical error (Table 6).

Table 6. Results of simulations with Katja phantom for which different material compositions were tested. Results from scenarios 1, 2 and 3 of Group 1 are shown.

|  | Mother Katja Fetus Katja | Mother Katja Fetus UF | Mother UF Fetus UF | Variation between compositions (%) |
|---|---|---|---|---|
| Photon dose per target dose [nGy/Gy] | 169 | 169 | 170 | 0.3 % |
| Neutron dose equivalent per target dose [nSv/Gy] | 641 | 642 | 643 | 0.2 % |
| Total dose equivalent per target dose [nSv/Gy] | 810 | 811 | 813 | 0.2 % |

**Discussion**

Thanks to the reduced out-of-field doses associated to proton therapy, pregnant cancer patients could benefit from this technique. Nevertheless, calculation of fetal dose using Monte Carlo simulations are of key importance and require the use of computational pregnant phantoms. Moreover, challenges related to the creation of secondary neutrons need to be considered.

In the current work three different groups evaluated the total dose equivalent delivered to the fetus during proton radiotherapy of the brain. For this purpose, two different versions of the same code, MCNP, were used as well as three different phantoms representing the second part of the second trimester of the pregnancy.

Firstly, the SOBP shape for each participant and each phantom was checked by setting up the mesh tally in the head of each phantom and showed very good agreement between the groups.

Next, we compared fetal doses between groups for the same code. MCNPX shows slightly less variation within the results among the groups than MCNP6.2. Differences between the groups could be attributed to the slightly different definition of the beam parameters and positioning of the beam to the center of the brain. Interestingly when running the same input file in the 2 versions of MCNP we noted very good agreement (within 6%) for the total fetal doses which ensures there is no difference in using the 2 versions of MCNP. Nevertheless, we would like to highlight the potential pitfalls that could arise when not taking enough care in the creation of the input file. We noted in setting up the simulations that atomic number of the element (without mass number), which is anyway ignored for photons and electrons, is an important identifier to select appropriate neutron cross section libraries for a specific element. We did not report on these results in the current paper, but it should be mentioned that the error lead to an underestimation of the dose up to 50% for MCNP6.2. For MCNPX the omit of the mass number was less important and only resulted in differences up to 10% related to the different cross section libraries selected by default between the 2 MCNP version. These observations encouraged us to specify the atomic number of the element, mass number of the nuclide and cross-section identifier for each material and all reported simulated data are from the specific settings, which can be observed in the supplemental table 2.

The fetal dose equivalent calculated for the Katja phantom was around 0.8 µSv/Gy and was dominated by neutrons (81%). When compared to published simulated data for pencil beam scanning proton therapy of the brain during pregnancy (Geng et al., 2015) our results are up to 3 times lower which could be attributed to different target volumes which was 4 times lower in our study (47 cm$^3$) compared to Geng et al (190 cm$^3$) as well as the use of different proton energies.

Unfortunately the majority of published data (both experimental and simulated) for proton therapy are for the passive systems, which does not allow comparison to our simulated data. A comparison can be made to the experimental data presented by Mesoloras (Mesoloras et al., 2006) who quantified the scattered neutron dose equivalent, in an experimental setup without scatter foils, range modulators and patient collimators. Results show fetus doses of 5.5 µSv/Gy at 13.4 cm to the field edge. The beam energy was up to 157 MeV, and the phantom was set up to be in the second trimester of pregnancy. They further investigated how the dose drops when the distance from the field edge increases, and found a drop of one order of magnitude at a distance 50 cm from the field edge. Since in our simulations fetus is at distance even further than 50 cm from the field edge, these experimental result could be considered in line with our simulation results.

The UF phantoms from both gestational ages have significantly lower dose equivalent to the whole fetus than the Katja phantom (approximately 46 %), which was surprising. First we verified the impact on difference in the tissue composition between the phantoms (Katja versus UF), nevertheless the impact of different tissue compositions was small as results agreed within 0.3 %, which is below the statistical error. Therefore, the difference in tissue composition is not responsible for the large discrepancy in the final result between UF and Katja phantom. Next we looked into geometrical differences between the phantoms and provided good overview for the geometrical characteristics in Figure 1 and 3. The phantom representation of the Katja is thinner so there is less mass to attenuate the stray radiation. Katja has slightly tilted head so positioning of the beam is also different relatively to the fetus, since there is more air between brain and belly. In other words, Katjas belly is directly exposed to the scattered neutrons, and they pass through air, so there is no tissue to attenuate and moderate the neutrons. As such Katja has elevated neutron flux in the top of the belly which causes the fetus to receive higher dose from neutrons.

Moreover, the fetus positioning in the belly is different for each phantom. The distance from the field edge to the center of the fetus was 55 and 64 cm for the Katja and UF 25 respectively. As described by Mesoloras (Mesoloras et al., 2006) a slight increase in the distance to the field can reduce the fetus neutron dose.
Our results suggest that dose estimation to the fetus require integration of as much as possible patients' and fetus anatomy when modeling fetus dose in clinical settings. Nevertheless, during pregnancy the fetus is in constant movement, and also moves during the radiotherapy treatment of the mother. As such its position and orientation will differ from fraction to fraction and introduce uncertainties on the fetal dose calculations.

In the near future our intention is to move to the more clinical case, where the realistic radiotherapy plans will be applied to the DICOM-CT images of the UF phantom. In this way, we will be able to assess total dose to the fetus involving clinical radiotherapy plans and allow selection of treatment parameters for optimal radiation protection of the fetus. Moreover, we would like to perform validation measurements with physical phantoms in proton therapy to benchmark our calculations. Within EURADOS WG9 we have well-characterized detector systems to measure this mixed field of radiation and initiatives have been launched to create physical pregnant phantoms, including the use of 3D printing technology.

According to the authors experience with voxelized phantoms several issues can be encountered and the following advice can be given. High number of voxels requires high computer performance for using the MCNP geometry plotter and for preforming simulations in general. Usage of DBCN card in the code, in order to skip the geometry check, speeded up the calculations significantly, especially for the UF phantoms which have high voxel number. Some of the participants had problems with running simulations with such a large voxel numbers, and the simulations lasted for several days (even weeks). Some versions of the MCNP were not even able to start the simulations at all.

## Conclusions

Results obtained by three different groups differ up to 8 and 12 % for MCNPX and MCNP6.2 respectively, which may be attributed to slightly different definition of the simulation parameters by each participant.

Following a brain proton therapy irradiation, the fetus dose equivalent on different phantoms determined to be between 0.4 and 0.8 µSv/Gy which is in correspondence with the previously published simulated and experimental data.

Result obtained by two different phantoms differ by factor of 2 which could not be attributed to minor differences in tissue composition between phantoms. According to the conducted simulations reason for this discrepancy comes from the positioning of the fetus and the geometrical characteristics of the mother. These results highlight the uncertainty related to fetal dose calculations due to varying positioning and orientation of the fetus between patients and during the course of treatment and emphasized the need to integrate as much as possible female and fetus anatomy when modeling fetus dose in a clinical settings.

## Acknowledgments

European radiation dosimetry group (EURADOS) is a non-profit association that engages research in various fields of radiation dosimetry. Working group 9 (WG9) is mainly dedicated to estimation of out–of-field doses during radiotherapy and recently planned to include out-of-field dose estimations and calculations of pregnant patients undergoing radiotherapy. Working group 6 (WG6) is dedicated to the issues regarding the computational dosimetry and Monte Carlo simulations. Both WG6 and WG9 members participated in this task. Groups that have participated in this exercise come from Belgian Nuclear Research Centre (SCK CEN), National Centre for Nuclear Research in Poland and Faculty of medicine in Osijek, Croatia.

The authors acknowledge Groups from University of Florida and Helmholtz University Zentum Münich for providing us the computational phantoms required for the calculations.

Michał Kuć from National Centre for Nuclear Research participated in phantom preparation prior the simulations.

Supplemental Table 1. Overview of different proton energies, weights and sigma values of the SOBP modelled in this work

| Energy [MeV] | Weight | Sigma [MeV] |
|---|---|---|
| 116.27 | 0.307 | 1.746 |
| 114.19 | 0.096 | 1.714 |
| 112.16 | 0.076 | 1.684 |
| 110.16 | 0.061 | 1.654 |
| 108.21 | 0.052 | 1.625 |
| 106.28 | 0.045 | 1.596 |
| 104.38 | 0.039 | 1.567 |
| 102.51 | 0.035 | 1.539 |
| 100.65 | 0.031 | 1.511 |
| 98.82 | 0.028 | 1.484 |
| 96.99 | 0.026 | 1.456 |
| 95.17 | 0.024 | 1.429 |
| 93.36 | 0.023 | 1.402 |
| 91.54 | 0.022 | 1.374 |
| 89.73 | 0.021 | 1.347 |
| 87.91 | 0.02 | 1.32 |
| 86.08 | 0.019 | 1.292 |
| 84.24 | 0.019 | 1.265 |
| 82.37 | 0.019 | 1.237 |
| 80.49 | 0.018 | 1.208 |
| 78.58 | 0.020 | 1.180 |

Supplemental Table 2. Overview of element definitions (ZAID), cross section libraries and models used in each simulation. *c* denotes Continuous-energy neutron data libraries, *u* denotes Continuous-energy photonuclear data libraries, *p* denotes Continuous-energy photoatomic data libraries and *h* proton data libraries, *m* represents usage of the models.

**MCNPx**

| | MCNPX_Institute 1 | | | | MCNPX_Institute 2 | | | | MCNPX_Institute 3 | | | | |
|---|---|---|---|---|---|---|---|---|---|---|---|---|---|
| El. | ZAID | c | u | p | h | ZAID | c | u | p | h | ZAID | c | u | p | h |
| H | 1001 | 80c | m | 84p | 70h | 1001 | 80c | m | 84p | 70h | 1001 | 80c | m | 84p | 70h |
| C | 6000 | 80c | 24u | 84p | 70h | 6000 | 80c | 24u | 84p | 70h | 6000/6012 | 80c | m/24u | 84p | 70h |
| N | 7014 | 80c | 70u | 84p | 70h | 7014 | 80c | 70u | 84p | 70h | 7014 | 80c | 70u | 84p | 70h |
| O | 8016 | 80c | 24u | 84p | 70h | 8016 | 80c | m | 84p | 70h | 8016 | 80c | 24u | 84p | 70h |
| Na | 11023 | 80c | 70u | 84p | m | 11023 | 80c | 70u | 84p | m | 11023 | 80c | 70u | 84p | m |
| Mg | 12000 | 62c | m | 84p | m | 12000 | 62c | m | 84p | m | 12000 | 62c | m | 84p | m |
| P | 15031 | 80c | m | 84p | 70h | 15031 | 80c | m | 84p | 70h | 15031 | 80c | m | 84p | 70h |
| S | 16000 | 62c | m | 84p | m | 16000 | 62c | m | 84p | m | 16000 | 62c | m | 84p | m |
| Cl | 17000 | 66c | m | 84p | m | 17000 | 66c | m | 84p | m | 17000 | 66c | m | 84p | m |
| Ar | 18040 | 80c | 70u | 84p | m | 18040 | 80c | 70u | 84p | m | * | | | | |
| K | 19000 | 62c | m | 84p | m | 19000 | 62c | m | 84p | m | 19000 | 62c | m | 84p | m |
| Ca | 20040 | 80c | 24u | 84p | 70h | 20040 | 80c | m | 84p | 70h | 20000/20040 | 62c | m | 84p | m |
| Fe | 26056 | 80c | 24u | 84p | 70h | 26056 | 80c | 24u | 84p | 70h | 26056 | 62c/80c | m/24u | 84p | 70h |
| I | 53127 | 80c | 70u | 84p | m | 53127 | 80c | 70u | 84p | m | 53127 | 80c | 24u | 84p | 70h |
| | | | | | | | | | | | 1001 | 80c | 70u | 84p | m |

**MCNP6.2**

| | MCNP6.2_Institute 1 | | | | MCNP6.2_Institute 2 | | | | MCNP6.2_Institute 3 | | | | |
|---|---|---|---|---|---|---|---|---|---|---|---|---|---|
| El. | ZAID | c | u | p | h | ZAID | c | u | p | h | ZAID | c | u | p | h |
| H | 1001 | 02c | 24u | 84p | 70h | 1001 | 80c | m | 84p | 70h | 1001 | 90c | m | 84p | 70h |
| C | 6000 | 80c | 24u | 84p | 70h | 6000 | 80c | 24u | 84p | 70h | 6000/6012 | 80c | m/24u | 84p | m/70h |
| N | 7014 | 02c | 24u | 84p | 70h | 7014 | 80c | 70u | 84p | 70h | 7014 | 80c | 70u | 84p | 70h |
| O | 8016 | 02c | 24u | 84p | 70h | 8016 | 80c | 24u | 84p | 70h | 8016 | 80c | 24u | 84p | 70h |
| Na | 11023 | 02c | 24u | 84p | 70h | 11023 | 80c | 70u | 84p | m | 11023 | 80c | 70u | 84p | m |
| Mg | 12000 | 62c | 24u | 84p | 70h | 12000 | 62c | m | 84p | m | 12000 | 62c | m | 84p | m |
| P | 15031 | 02c | 24u | 84p | 70h | 15031 | 80c | m | 84p | 70h | 15031 | 80c | m | 84p | 70h |
| S | 16000 | 62c | 24u | 84p | m | 16000 | 62c | m | 84p | m | 16000 | 62c | m | 84p | m |
| Cl | 17000 | 66c | 24u | 84p | m | 17000 | 66c | m | 84p | m | 17000 | 66c | m | 84p | m |
| Ar | 18040 | 80c | 24u | 84p | 70h | 18040 | 80c | 70u | 84p | m | * | | | | |
| K | 19000 | 62c | 24u | 84p | m | 19000 | 62c | m | 84p | m | 19000 | 62c | m | 84p | m |
| Ca | 20040 | 02c | 24u | 84p | 70h | 20040 | 80c | 24u | 84p | 70h | 20000/20040 | 80c | m/24u | 84p | m/70h |
| Fe | 26056 | 02c | 24u | 84p | 70h | 26056 | 80c | 24u | 84p | 70h | 26056 | 80c | 24u | 84p | 70h |

I   53127  02c  $^{24}_u$  $^{84}_p$  70h  53127  80c  $^{70}_u$  84p  m        53127        80c        70u  84p    m

* Group 3 did not used Argon in air definition